\documentclass[12pt]{article}

\def\centereps#1#2#3{\centerline{\includegraphics[width=#1, height=#2]{#3}}}
\def\be{\begin{equation}} 
\def\ee{\end{equation}}

\usepackage{epsfig,graphicx}
\begin{document}
\begin{center}
{\huge \bf The Electrochemical Reduction of Hydrogen in the Presence of Bisulfate on Platinum(111)
}\\ 
\vspace{0.2in}

{\it L. Blum,
Department of Physics, P.O. Box 23343, 
University of Puerto  Rico, 
Rio Piedras, PR 00931-3343}\\

and \\
{\it Dale A. Huckaby,
Department of Chemistry, 
Texas Christian University,  
Fort Worth, TX 76129
}\\
\vspace{0.1in}

{ \Large Abstract}\\
\end{center}
                                                                                                                                            
 A new model for the intermediate compound of the hydrogen evolution reaction (HER) is proposed,  for the  electrochemical reduction of hydrogen in the presence of bisulfate on platinum(111). The formation of this compound, a regular 2 dimensional honeycomb ice lattice, occurs by a first order phase transitions that involves the reorientation of water molecules. The model is analyzed using  new and  simple effective cluster approach which highlights the relevant transitions in system. This method is based on the cluster variation method used successfully in our previous work on the UPD of Cu onto 
Au(111), and permits us to explore a large region of parameter space, an essential feature to study this complex system. The theory makes full use of the properties of the diffuse layer:  The water molecule is  reoriented as the potential is changed.  For positive potentials it forms linear chains which are responsible for the $\sqrt 3 \times \sqrt 7$ structure of the sulfate observed by STM. At negative potentials water turns so that its dipole points towards the Pt. Then it will form a regular honeycomb  network of hydrogen 
bonded molecules, with  the sulfate at the center of the hexagons. Then the bisulfate is desorbed, leaving the honeycomb HER structure behind. Our model thus provides an  explanation of the well known fact that only 2/3 of the Pt atoms participate in the electroreduction of the hydrogen. The  theory implies geometrical constraints to  the water potential: It should\\

$\bullet$ be of tetrahedral coordination.\\

$\bullet$ be analytical to be able to include the double layer effects, the ions and external fields, in the calculations.\\

$\bullet$ The angular part of the potential must be relatively soft, since the angle O-H-O of
the hydrogen bond will be bent to keep the water near the metal surface.\\
The analytical tetrahedral Yukagua model of water satisfies all of these requirements. \\

\section { Introduction}

The underpotential deposition of hydrogen on platinum(111) in aqueous sulfuric acid was studied by Clavilier et al. using his pioneering techniques of preparing monocrystalline electrode
 surfaces\cite{clav1,clav2}.\\

In Fig.1 \ref{one} we show a recent voltammogram for this system obtained by V. Climent \cite{ortsvolt} on a rather perfect single crystal. For our discussion we will divide this voltammogram in three regions:
\begin{figure}
\parbox{4.5in}{\centereps{16cm}{16cm}{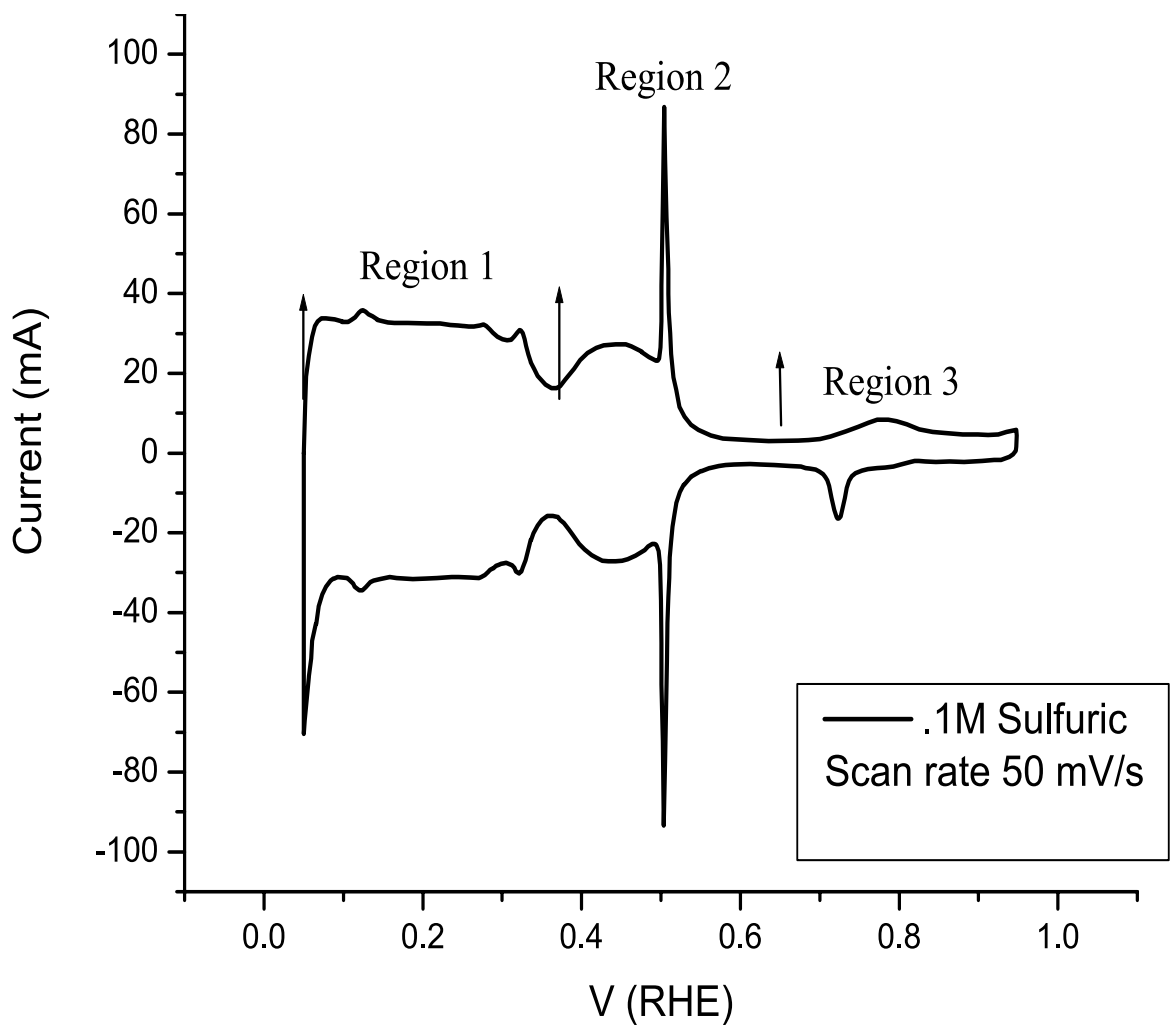}
\caption{The voltammogram for the $H_2/Pt(111)/SO_4H_2$ (courtesy of V. Climent)
system  is divided in three regions: 1)The UPD
 region, 2) the CP butterfly region and 3) the positive region}}
\label{one}
\end{figure}
\begin{enumerate}
\item The H-UPD region: A new structure of the HER intermediate is proposed, which explains the observed 2/3 of a monolayer yield for this surface. 

\item the CP butterfly region: Here there are two distinct features, a sharp spike, which we identify with   the {\it formation} of a $\sqrt 3 \times \sqrt 3$ lattice of water-like molecules,  with its dipole pointing {\it up}.

\item the positive region: In this region the $\sqrt 3 \times \sqrt 7$ structure is formed by 'polywater' chains  of water molecules with its dipole pointing {\it down}
\end{enumerate}

We note that these steps are  similar to the ones proposed by us  for the UPD of Cu/Au(111)\cite{hb1,hb2,hb3}: In reverse order:
 \begin{itemize}
\item Region 3: The formation of stacked, elongated hexagonal structures corresponding \cite{itaya95,itaya00,soriga00} to water-bisulfate coadsorption is explained by the formation of a hydrogen bonded water chain\cite{orts1,orts2}. There is no phase transition in this case.
\item Region 2: The spike here is  due to the co-adsorption of bisulfate and water, forming in fact the same honeycomb structure  as in the Cu/Au(111) case. Now the sulfate simply leaves an empty site, while in the copper case it was replaced by a copper via a phase transition.
\item Region 1:
In this region the transition is different because as the sulfate is desorbed, its site was filled by another atom of copper. In our present case, the hydrogen bond of the hexagons are saturated, and the sulfate is simply desorbed.
\end{itemize}

In section 2 we outline the basic formalism and our method of computing the fugacities of our active compounds. In section 3 we discuss the positive Region 3 of the voltammogram. In section 4 we present our model for the CP butterfly region, and finally in section 5 we discuss our model HER intermediate. In the final section we compare our resulting voltammogram to experiment.\\

\section{BASIC FORMALISM}

We use a microsopic model, based on the sticky site model (SSM) \cite{rosi1,dah1,dah2}, in which the exact hamiltonian of the entire interface is projected onto an equivalent 'electrode' surface lattice: This means that all solvation and double layer  effects are included. For example, for dilute solutions such as the one of our present example the Gouy-Chapman approximation is quite accurate, as has been shown in previous work \cite{hb3}.\\

One of the problems in surface electrochemistry is the complexity of even seemingly simple systems, like the one of this paper. Even with modern day resources it is very difficult, if not outright impossible, to explore the parameter space of this problem by computer simulation. For this reason we have used  the 'Cluster Variation Method (CVM)\cite{gug1,bell1,shin1,hshin,shin2,hlb1} to calculate the adsorption isotherms and generated a voltammogram  for the deposition of H on Pt(111) in the presence of 0.1 M aqueous sulfuric acid. In our  our previous work on the  UPD of Cu on Au(111)\cite{hb3,hb1,hb2,hlb1} we used  the "cluster approximation"\cite{hlb1}. This method  successfully predicted the local structure  and transformations of the voltage dependent phases that were later observed.
For the UPD of hydrogen and the bisulfate anion we use 
first a simplified version of this method, the PCVM ( PADE-CVM) approximation which further simplifies the calculations. This work will be described in a separate paper \cite{BHL01.2}
 In our present work we will use a still simpler theory that we call the assymptotic cluster model (ACM), where the behavior of  the model is described by a few assymptotically large diagrams which determine its transitions. The interaction  parameters are obtained from theory or simulation.\\

The adsorption of  species $i$ on the platinum surface primarily depend on the activities of these adsorbates and the lateral interactions existing between them\cite{rosi1,dah1,dah2}. The activity of a species $i$ is given by the product between a sticky parameter $\lambda_i(\psi)$ and the species contact density at the surface $\rho^\circ_i(0,\psi)$,
 as
\be
z_i=\lambda_i(\psi) \rho_i^\circ (0,\psi) \mbox{  .}
\label{zor}
\ee
The sticky parameter $\lambda_i(\psi)$ is a phenomenological parameter characterizing the quantum mechanical interaction between the adsorbate and surface, and will in general, be potential dependent.  We write \cite{hlb1}\\

\be
\lambda_i(\psi)= e^{-\beta \kappa_i \zeta_i e(\psi - \psi_{ref})} \mbox{  ,}
\ee
where $e$ is the elementary charge, $\zeta_i$ is the partial charge of the adsorbate at the surface, and $\kappa_i$ is a binding constant characterizing the overlap of electron orbitals between the adsorbate and surface atoms.
The potential dependence of the contact density is given to a very good approximation by the Gouy-Chapman expression
\be
\rho_i^\circ(0,\psi)=\rho_i^\circ(0,0)e^{-\nu_i e \beta (\psi-\psi_{pzc})}\mbox{  ,}
\ee
where $\rho_i^\circ(0,\psi_{pzc})$ is the density at a smooth (i.e. without sticky sites) surface\cite{dah2} with a potential of zero charge, and $\nu_i$ is the ionic charge of the adsorbate. It has been verified in previous work \cite{hb3} that for our relatively dilute concentrations of sulfate ions this is a quite good approximation.\\

The potential dependence of the activity $z_i$ is then found to be
\be
z_i=\lambda_i (0) \rho_i^\circ(0,0)e^{-\beta \gamma_i e(\psi-\psi_{0})} \mbox{  ,}
\ee
where $\gamma_i=\nu_i+\kappa_i \zeta_i$ is an effective electrovalence of species $i$.\\

We  obtain analytical expressions representing clusters of interacting adsorbates which are  used to calculate the coverages of hydrogen and the bisulfate anion as a function of potential.  The crucial parameters in our present theory are the water orientation and the bisulfate adsorption.

\subsection{
THE ADSORPTION   AND ORIENTATION OF WATER AND (BI)SULFATE}

We discuss first the potential dependent adsorption fugacities of bisulfate and water. In our model water is adsorbed on the atop position  when its dipole is pointing down and in the hollow site when the dipole is up. \\

\subsubsection{BISULFATE}

The 
inner layer equivalent fugacity ${z}_S$ for the adsorption of the bisulfate is
\be 
{z}_S=\lambda_S^0\rho_S^0(0,0)e^{-\zeta_S \beta e (\psi-\psi_{S}^{Re})}
\label{zt3a}                
\ee
where  $\beta=1/kT$ is the Boltzmann thermal factor, 
the electrosorption valency of the bisulfate is  $\zeta_S=-1$, 
and  $\psi_{S}^{Re}$ is the electrosorption  reference potential, that depends on the nature of the substrate.
The sticking coefficient
 can be interpreted as $\lambda_S (\psi)=\exp{[\beta \mu_S]}$, 
 with  $\mu_S$ as the  free energy change that occurs when a bisulfate ion
binds to the metal surface. 
$\rho_S^0(0,\psi)$ is the inner layer local density of bisulfate 
for a local potential $\psi$, which is estimated from the Gouy-Chapman formula.
\be
{z}_S=\lambda_S^0\rho_S^0(0,0)e^{\beta e (\psi-\psi_{S}^{Me})}
={z}_S^0 e^{\beta e (\psi-\psi_{S}^{pzc})}
\label{muv1}
\ee 
where we take
\be
\psi_ {S}^{pzc}= 0.8 V (RHE)
\label{muv12}
\ee
and  the bisulfate fugacity $z_S$ is ($z_S^0=1, \qquad T=298.16 K$)
\be
z_S= e^{38.922 [\psi-\psi_ {S}^{pzc}]},
\label{zuv1}
\ee
The electrovalence is taken as $\zeta_S=-1$.\\

\subsubsection{WATER}

For positive electrodes only the oxygen of the water molecule binds to the electrode metal atoms, since  the strong  electric field ${\mathbf E}$ will orient  the large dipole $ {\mathbf \mu}_W$ of the water molecule. The preferred position at positive potential bias is that of an inverted tripod, with two of the hydrogens 
pointing upwards. The adsorption fugacity of the 'oriented' water can be written as
\be
{z}_W= \rho_W \lambda_W^0  g_d^W
\label{owt}
\ee
where$ \rho_W=3.345 \AA^{-3}$ is the density of water and the adsorption parameter $\lambda_W^0$ is
\be
\lambda_W^0= e^{\beta \mu_W  E_z(\Delta \psi_W)}\qquad \Delta \psi_W\equiv (\psi-\psi_{W}^{Me})
\label{wuv1}
\ee 

\begin{figure}
\centereps{10cm}{7cm}{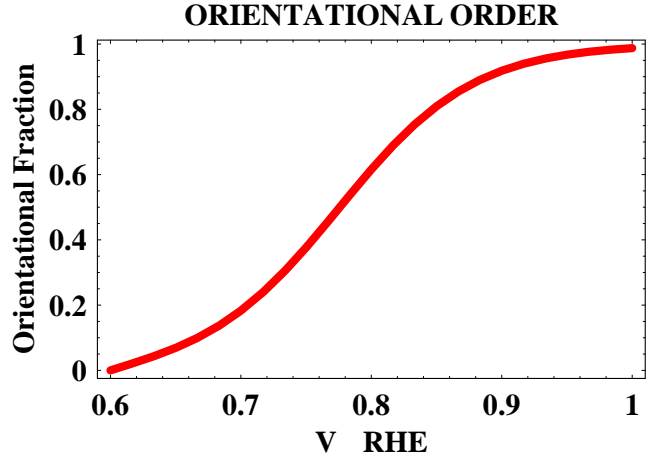}
\caption{Orientational order parameter.}
\label{two}
\end{figure}

\begin{figure}
\parbox{4.5in}{\centereps{10cm}{7cm}{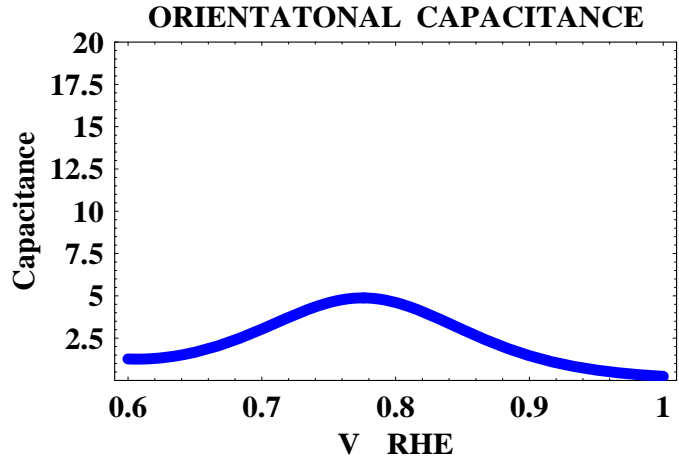}
\caption{Orientational contribution to the capacitance / voltammogram.}}
\label{three}
\end{figure}

Here $g_d^W$ is the orientation parameter of the water molecules in contact with the electrode, $\rho_W $ is the bulk density of water. We use the Gouy-Chapman theory to get the perpendicular component of the electric field at contact with the electrode
\be
  E_z(\Delta \psi_W)= (-) \frac{ 2 \kappa }{\beta e}\sinh (\Delta \psi_W/2)\qquad where \qquad \Delta\psi_W=38.94 \Delta\phi(Volts)
\label{gc1}
\ee
 A  simple estimate of the orientation parameter of the water molecules near a charged electrode,$g_d^W$ can be obtained using a suitably adapted form of the mean spherical model \cite{bhid1,blve,blsc}:
\be 
g_d^W=-\frac{\sqrt{3}\beta \mu_W E_z(\Delta \psi_W)}{\lambda(2+\lambda) }  [ 1 -1/8(\lambda+1)\Gamma \sigma  ]
\label{gc1a}
\ee
To conform to the high coupling- low concentration limit we use the EXP approximation

\be 
g_d^W=-\frac{\sqrt{3}\beta \mu_W E_z(\Delta \psi_W) }{\lambda(2+\lambda) }  \exp{\{ -(1/4)(\lambda+1)\Gamma \sigma \} }
\label{gc2}
\ee
The parameters in this equation are $\kappa$, the Debye Hueckel screening length,$\Gamma$ is the MSA screening length, $\mu_W$, the dipole moment of water and $\sigma$, the diameter of the water molecule (2.8 angstrom). The polarization parameter $\lambda$ is the MSA Wertheim parameter, obtained from
\be
4\sqrt{\epsilon_W}=\lambda (1+\lambda)^2
\ee
where $\epsilon_W$ is the dielectric constant of water, (78.4 at 298 $^o$K), for which we get 
\[
\lambda=2.652
\]
From these we  get for the dipole model
\be
g_d^W=\frac{2\sqrt{3}\kappa \mu_W  }{e \lambda(2+\lambda) } \sinh{(\Delta \psi_W/2)} \exp{\{ -(1/4)(\lambda+1)\Gamma\sigma \} }
\label{gc3}
\ee
or
\be
g_d^W=0.0943 \sinh{(19.46\Delta \phi_W)} \exp{\{ -2.4635\Gamma\sigma \} }
\label{gor}
\ee
which is used in conjunction with  Eq.(\ref{zor}) to compute the average orientation opf the water molecules. The results of this calculation are shown in Fig. 2(\ref{two}), and the contribution to the differential capacitance (in region 3) are shown in Fig.3(\ref{three}). 

\begin{figure}
\parbox{4.5in}{\centereps{10cm}{10cm}{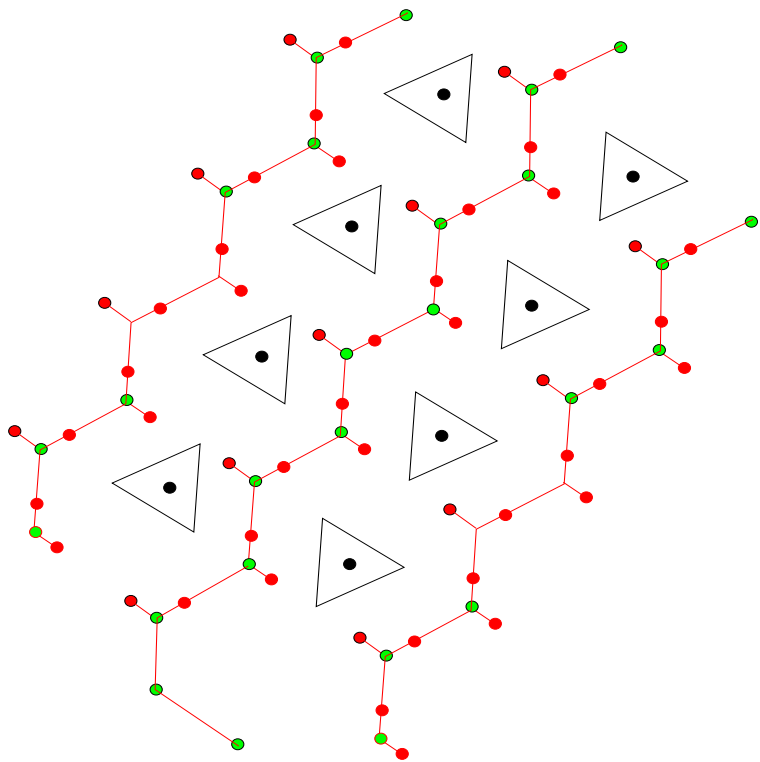}
\caption{Water chain structure.The triangles are bisulfate ions. The hydrogens on the hydrogen bonds are gray dots.}}
\label{four}
\end{figure}

 This procedure will be extended to include the full 
Yukagua model \cite{bvede} in the future. We do not expect however large differences since the main term in the re-orientation hamiltonian is the dipolar one, and the predicted capacitance for region 3, obtained from this model agrees reasonably well with experiment. 
The  chain structure is shown in Fig.4(\ref{four}).

\section{REGION 3: THE WATER CHAIN REGION}
The STM structures seen in these systems in the positive end of the voltammogram by various authors \cite{ito1,ito2,futnik}, consists of elongated $\sqrt{3}\times \sqrt{7}$ hexagons rather than regular hexagons. The elongated structures correspond to water coadsorption as was shown by Ito. More recent  high resolution STM experiments \cite{itaya00,soriga00} have been interpreted as being the result of hydrogen bonded water chains. However, the structures that are proposed are unstable in the presence of the high field gradients  \cite{mott61} of the electric double layer, since the dipoles are alternated in order to keep the hydrogen bond straight. A different model is proposed in this paper in which all dipoles are polarized the same way. However this requires bending of the hydrogen bond. Our point here is that this is actually the case, even in normal ice Ih \cite{chidam}, and very clearly in liquid water \cite{bvede}. This model also explains why when the polarity is changed, those chains no longer exist and the regular  $\sqrt{3}\times \sqrt{3}$  hexagons are observed \cite{hb1,hb3}. The details of this model, as well as its consequences will be discussed in a separate paper \cite{orts1,orts2}

The formation of water hydrogen bonded chains occurs because the properly oriented water molecules form hydrogen bonds with the lone pair and one of the two hydrogens of the 'upper' plane. In principle any of the two  can form the next bond, and therefore the 'water chain' chain will consist of units forming an angle 
of 142$^o$ to 152$^o$ instead of the straight 180$^o$, and the 'free' hydrogen will be pointing to either side of the chain. If consecutive free hydrogens 
point to opposite sides of the chain, then the space between next nearest neighbors can be occupied by a (bi)sulfate ion. If the nearest neighbors free hydrogens 
point to the same side, then the chain will bend, and there will be a steric hindrance for the (bi)sulfate ion 
in the neighboring site. 
The model can be mapped into an equivalent one dimensional polymer with internal degrees of freedom. As such no first order transition occurs in the formation of these chains of elongatged hexagons.  A detailed discussion of this model is left for a future publication\cite{orts2}.

\subsection {Contribution to the voltammogram}
The main contributor to the current is  the capacitive current  due  to the flipping of the water molecules, which  is computed from the orientational fugacity Eqs.(\ref{zor},\ref{gor})using a Langmiur type equation
\be
\theta_W\simeq \frac{1}{5} \left[\frac{2 \alpha z_W e^{K_S}}{1+ \alpha z_W e^{K_S}}\right]
\ee
where $\alpha$ is a constant that determines the position of the maximum of the voltammogram in region 3, and has been adjusted for the time being.
The change in the orientation of the water molecules will produce a change in the capacitive current. 
\be
j_C= (1/A) \left[C  +\psi  \frac{d C}{d\psi}\right]
\frac{d \psi}{dt} \propto (M/A) e \zeta\frac{d \theta }{d\psi}
\frac{d \psi}{dt} 
\ee

\noindent where $C$ is the integral capacitance, $M/A$ is the number of adsorption
sites per area, e is the
elementary charge, and $\zeta$ is the partial charge of the
adsorbate.\\

\section{REGION 2: THE CP BUTTERFLY }

The classic shape of the cyclic voltammogram of the hydrogen UPD on a platinum single crystal with (111) orientation has been likened to the silhouette of a butterfly, hence this particular CV profile is known as Clavilier's {\it papillon}(CP)\cite{clav1}. This system has been studied experimentally by a large number of authors\cite{clav2,furuya,ross1,ito1,ito2,tadj1,nishi,feliu1,feliu2,feliu3,jerk1,jerk2}\\

\begin{figure}
\parbox{4.5in}{\centereps{10cm}{5cm}{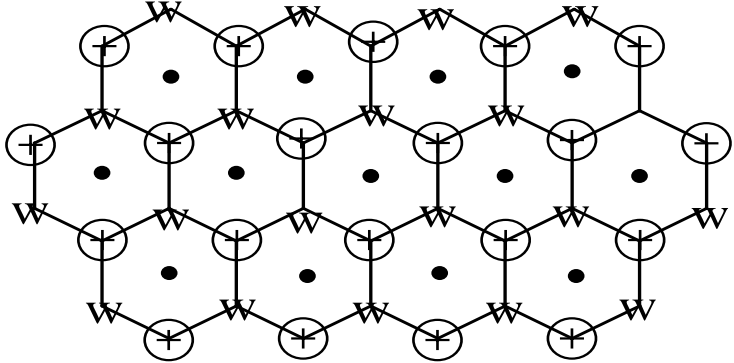}
\caption{The honeycomb bisulfate-water-hydronium lattice. It is commensurate with the {\it adsorption sites} of the Pt(111) face. The $\oplus$ corresponds to a hydronium,
 the $W$ to a  water with the $H$ pointing downwards, and the points are either an empty site or a bisulfate ion.}}
\label{five}
\end{figure}

The main point of our theory is the formation of an hexagonal two dimensional honeycomb two dimensional 'ice' structure ( the HER( Hydrogen Evolution Intermediate) structure shown in Fig.5(\ref{five}) that remains in place after the desorption of the bisulfate ion. The spike of the Clavilier papillon (CP, region 2)corresponds to the formation of an ordered, $\sqrt 3 \times \sqrt 3$, commensurate phase
\be
\frac{1}{3} SO_4H^- +\frac{1}{3}H_2O+ \frac{1}{3}H_3O^+
\rightarrow Lattice.
\ee
This is shown in Fig.5 
 The 'wing' of the CP \cite{kuni1,faguy1,wieck01}is associated with the desorption of (bi)sulfate, and
corresponds also in our theory to the desorption of the bisulfate. However in our case the HER intermediate, the water-hydronium  honeycomb (WHH)lattice stays intact, and is well represented by a Langmuir adsorption isotherm. This  desorption mechanism is different from the hard hexagon order-disorder transition, recently proposed by Koper et al. \cite{koper00}, since in their case the desorption produces necessarily a disordered lattice.

The CO displacement experiments  of Feliu and co-workers \cite{feliu94,feliu00} 
show very clearly that  the 1/3 corresponding to the 'wing' of the butterfly is capacitive and is due to the desorption of the bisulfate, which is what we propose in our theory. \\ 
The detailed PCVM treatment will be published elsewhere
\cite{BHL01.2}. Some highlights are:
\begin{itemize}
\item Input parameters are the fugacity of 'up' water (W), the fugacity of 'up' hydronium (H) and of the sulfate (S). However the formation of the honeycomb structure requires an equimolecular mixture of W and H, so that for the cluster theory we only need an effective
adsorbate, which is one half of the W-H compound, the semi-hydronium (SH). In our model the assymptotic occupancy is 1/3 of W, 1/3 of H, or 2/3 for the SH complex, but the effective charge of the SH complex is only 1/2
\item the relevant parameters are the S-S interactions, the SH-SH=W-H interactions and the SH-S interactions.
\end {itemize}
 
\begin{figure}
\parbox{4.5in}{\centereps{10cm}{10cm}{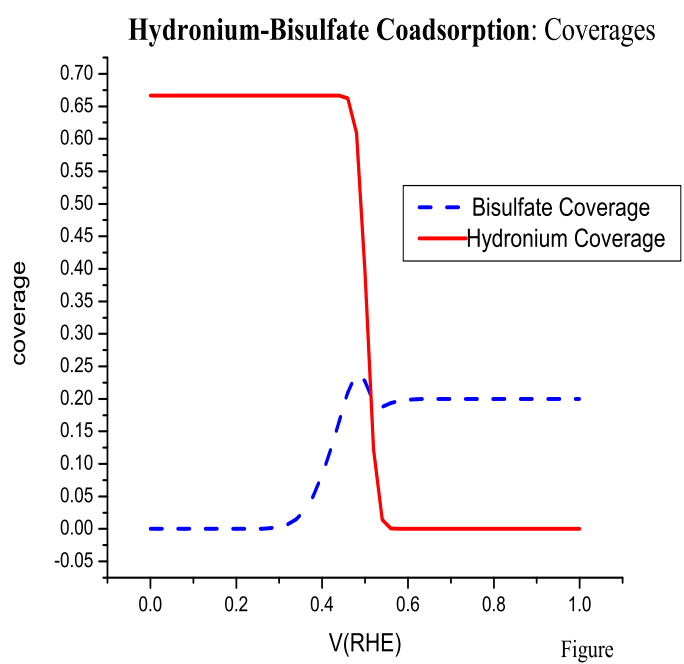}
\caption{Semihydronium-bisulfate coadsorption transition coverages.}}
\label{six}
\end{figure}

\begin{figure}
\parbox{4.5in}{\centereps{10cm}{10cm}{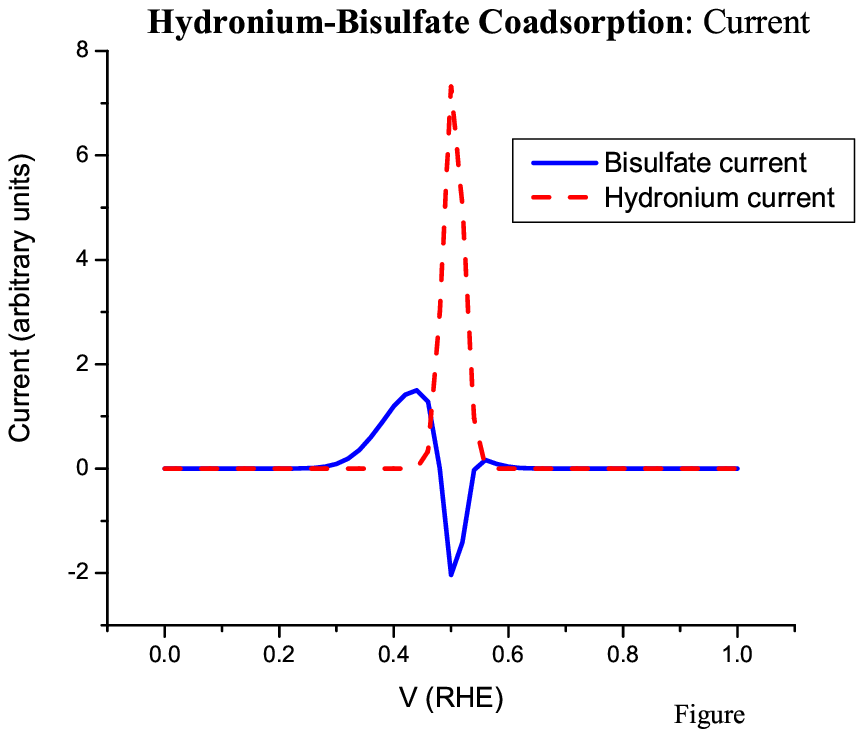}
\caption{Semihydronium-bisulfate coadsorption transition currents.}}
\label{seven}
\end{figure}

The result of the CVM are  a series of sharp and soft transitions\cite{hlb1}. In our case there  is only one sharp transition, which corresponds the co-adsorption of bisulfate and hydronium and  presumably, also the reorientation of water. The adsorption isotherms are represented by smoothened step functions:\\
For the bisulfate
\be
\theta_{S}=\frac{1}{6}\{1+ Erf[\Delta_S(\psi-\psi_S)]\}+\frac{1}{15}\{-1+ Erf[-\Delta_{SH}(\psi-\psi_{SH})]\}
\ee
and for the semihyronium complex
\be
\theta_{SH}=\frac{1}{3}\{1+ Erf[-\Delta_{SH}(\psi-\psi_{SH}))]\}
\ee
The positions and widths of the transitions are treated as adjustable parameters, which however correspond the well defined physical parameters: 
$\Delta_{SH}= 40$ is the semihydronium transition width;\\
$\psi_{SH}=0.5041$ is the position of the semihydronium transition, which occurs when there is simulataneous co-adsorption of bisulfate and semihydronium.\\
$\Delta_{S}= 12$ is the sulfate;\\
$\psi_{S}=0.44$ is the position of the sulfate transition, which is determined by the bisulfate-bisulfate repulsive interaction.\\
The bisulfate goes from  a coverage of 1/5 to 1/3 , and the semihydronium from 0 to 2/3. The exact values for the bisulfate coverages will depend on the other interaction parameters, and will be discussed in a more detailed PCVM analysis. The coverages for the semihydronium are fixed by the experimental hydrogen yield.
The results of the treatment for the adsorption isotherms  are shown in Figs.  6(\ref{six}) and 7(\ref{seven}). The agreement with the recent work of Kolics and Wieckowski\cite{wieck01} is qualitative: First the systems are not exactly the same, and in our case we predict a small glitch when the sulfate is re-adsorbed. We remark also that it is very likely that bisulfate and perchlorate co-adsorb \footnote{ Pechlorate is tetrahedral, the bond length is 1.325 $\AA$ in the crystal. Bisulfate cannot be measured experimentally.The in vacuum  structure has been calculated using state of the art DFT by S. Gaurei, is a slightly distorted tetrahedron with a S-O distance of 1.43 \AA. The major difference appears to be the dipole moment, which is zero for perchlorate and 3.80 D for bisulfate} , since they are isostructural and isoelectronic. This point will be discussed in the future. 
 
\section{THE UPD REGION 1: THE MODEL FOR THE HER INTERMEDIATE}
Region 1, which is called the H-UPD region \cite{bewick73}has been explored experimentally \cite{bewick81,tadj1}, and 
several proposals have been made for the HER intermediate
\cite{bewick82}, based on interpretations of the IR spectra. Our  model is based on the stochiometry and the analysis of the  phases of the voltammogram, the STM experiments and the recent radiotracer adsorption experiments, and is consistent with the experiment described by Peremans et al \cite{tadj1}
 Our HER model is  a  hydrogen bonded network of hexagonal rings, wwhich is a form of two dimensional ice. The chemical formula for a species is 
\be
[(H_5O_2)^+]_3
\ee
which forms a network of hexagonal rings. The explicit reaction taking place at the (111) face of the electrode is
\be
[(H_5O_2)^+]_3+ 6 e^-=3 H_2 + 3H_2O+ 3HO^-
\ee
where the 6 hydrogen ions that are neutralized are those
'trapped' in the {\it hollow} sites of the Pt(111) surface. This means that the spectra observed in the SFG experiments of Tadjeddine are likey to correspond to $Pt_3-H-O-H_{3/2}$ (semi-hydronium) complexes.
 The 'upd' hydrogen is really a 'semihydronium' ion, with the hydrogen stuck in the hollow site of the platinum lattice, and the dipole of the water molecule properly oriented. Recent quantum mechanical DFT calculations support this model \cite{car01}.

\begin{figure}
\centering
\parbox{4.5in}{\centereps{10cm}{10cm}{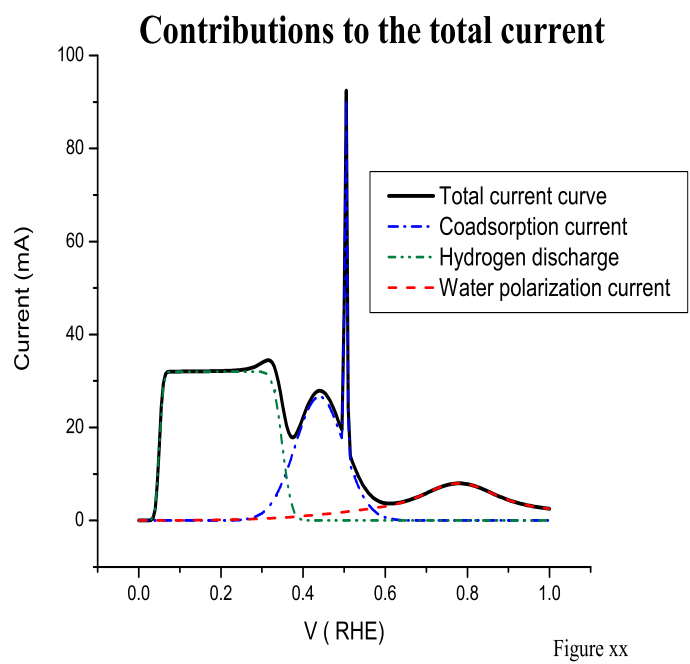}
\caption{Contributions to the  current.}}
\label{eight}
\end{figure}

\begin{figure}
\centering
\parbox{4.5in}{\centereps{12cm}{12cm}{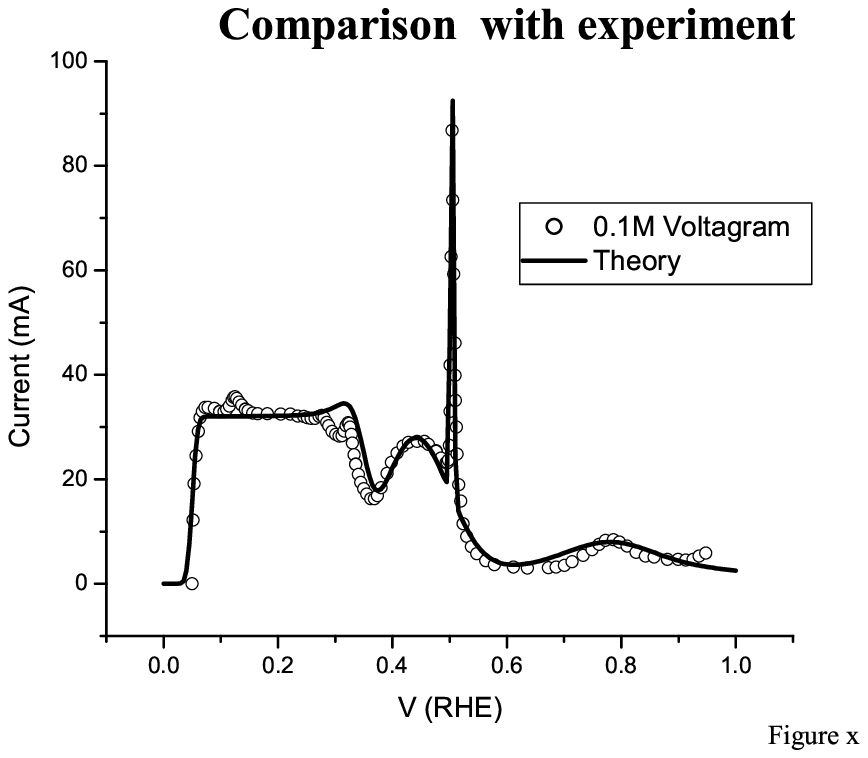}
\caption{Comparison between our model and the experimental voltammogram.The experimental data is courtesy of Dr. V. Climent}}
\label{nine}
\end{figure}

We assume that a 'surface state' is formed by the HER inermediate that discharges linearly with potential. Our model explains the 2/3  yield for the (111) face in a very natural way. The contribution of this region is only faradaic.

\be
j_F= (M/A) e \left[(\nu-\zeta)\frac{d \theta }{d\psi}- \theta   \frac{d \zeta }{d\psi}
\right]
\frac{d \psi}{dt} 
\ee

\noindent where  $M$ is the number of adsorption sites per area $A$,
e is the
elementary charge, $\nu$ is the electrovalence of the adsorbate in the bulk,
$\zeta$ is the partial charge of the adsorbate at the surface,
and $\psi$ the potential. ($\nu-\zeta$) is the charge per adsorbate transferred to the surface. Fig. 8(\ref{eight}) shows the contributions to the faradaic current due to the processes of adsorption and discharge of the hydrogen adsorbate. Fig. 9(\ref{nine}) shows the rather nice agreement with the experiment ( courtesy of Dr. V. Climent) for a 0.1M sulfuric acid solution. 

\section{ DISCUSSION OF RESULTS}

The main points in our work are that the the tetrahedral geometry of the water potential and the large dipole create two different water 'phases' for positively and negatively charged electrodes: Long chains at positive potentials and a honeycomb lattice for negative potentials. The formation of the honeycomb two dimensional ice phase is promoted by the bisulfate, or by other anions present, and the spike observed experimentally corresponds to this coadsorption. The honeycomb structure remains in place after the bisulfate has left, and is responsible for the 2/3 hydrogen yield of the Pt(111) face\\

\section{Acknowledgements}

This research was supported  by the
DOE-EPSCoR grant DE-FCO2-91ER75674, by The Robert A. Welch Foundation grant P-0446. The authors thank Profs. G. Jerkiewicz,  V. Climent,J.M. Orts and A. Wieckowski for the data files of  voltammograms, and Prof. R. Car and Dr. N. Marzari for useful discussions.

\bibliography{}

\end{document}